\documentclass[nofootinbib,notitlepage,superscriptaddress,aps,pra]{revtex4-1}

\usepackage[utf8x]{inputenc}
\usepackage{amsmath}
\usepackage{amssymb}

\newcommand{\su}{\mathfrak{su}}
\newcommand{\so}{\mathfrak{so}}

\newcommand{\C}{\mathbb{C}}
\newcommand{\eq}{\approx}

\newcommand{\Ad}{\text{Ad}}

\newcommand{\g}{\mathfrak{g}}
\newcommand{\ot}{\mathfrak{\otimes}}
\newcommand{\R}{{\cal R}}
\newcommand{\z}{z}
\newcommand{\zz}{z}

\begin{document}


\title{$\kappa$-de Sitter and $\kappa$-Poincaré symmetries emerging from
Chern-Simons (2+1)D gravity with a cosmological constant}
 
\author{Giacomo Rosati}
\email{giacomo.rosati@ca.infn.it}
\affiliation{INFN, Sezione di Cagliari\\ Cittadella Universitaria, 09042 Monserrato, Italy}

\begin{abstract} 
Defining a new r-matrix compatible with the scalar product at the basis of the Chern-Simons action for a particle coupled to (2+1) Lorentzian gravity with cosmological constant, I show how deformed symmetries of $\kappa$-de Sitter and, in the vanishing cosmological limit, of $\kappa$-Poincar\'e kind, arise naturally as quantum-deformation of three dimensional gravity.
I obtain moreover the non-commutative spacetime associated to these kinds of symmetries.
\end{abstract}


\maketitle

\section{Introduction}

The possibility for relativistic symmetries to be deformed at Planck scale ($1/M_p \sim 10^{-19} \text{GeV/c}^2$) has attracted much attention in the last 15 years particularly for its implications in quantum gravity phenomenology.
Models of quantum spacetime relying on this assumption have been indeed at the basis of the so-called DSR (Doubly Special Relativity, or, in a more general connotation, deformed relativistic symmetries) approach~\cite{gacdsr1IJMPD,joaoleePRDdsr,jurekDSRfirst} to quantum gravity phenomenology~\cite{AmelinoCamelia:2008qg},
whose main goal is to find experimental opportunities to test Planck scale modifications of relativistic laws of motion.
In this approach the relativistic symmetries are modified at the Planck scale in a way that preserves the relativity principle, in the sense that no preferred reference frame needs to be introduced~\cite{gacdsr1IJMPD}.

Even if still at its dawn, this research program has obtained several results especially concerning the opportunity to observe Planck scale signatures in the propagation of ultra-high energy particles from transient astrophysical sources~\cite{AmelinoCamelia:2008qg} (see~\cite{neutrini} for the most recent results),
where the cosmological distance travelled acts as amplifier for the tiny Planck-scale effects.
In this context the interplay between Planck-scale effects and the ones due to spacetime curvature/expansion of cosmological models becomes crucial~\cite{MarcianoInterplay,DSRdS,DSRFRW}.
This makes the construction and study of models of quantum spacetimes with non-vanishing cosmological constant at the top of quantum gravity phenomenology agenda.

$\kappa$-Poincaré symmetries~\cite{Lukierski:1991pn,MajidRuegg} and their generalization to $\kappa$-de Sitter\footnote{The $\kappa$-Poincaré Hopf algebra 
has been originally derived as a contraction of a specific $q$ deformation, in the Hopf algebraic sense, of the de Sitter group known as $q$-de Sitter~\cite{Lukierski:1991pn}. In this manuscript I refer to $\kappa$-de Sitter symmetries to indicate a generic class of $q$ deformations of de Sitter symmetries that reduce to $\kappa$-Poincaré in a suitable limit for vanishing cosmological constant.} provide one of the most interesting frameworks on which to realize these kinds of quantum spacetime models.
The mathematical formalism at the basis of these constructions is the one of the Hopf algebraic approach to quantum groups~(see for instance~~\cite{MajidFound}), 
which allows a description on the same footing of the deformed symmetry group and its (deformed) algebra.
It is relevant for the arguments presented in this manuscript to notice that these structures are characterized by a ``quantum duality principle'' establishing a correspondence between quantum universal enveloping algebras (i.e. Hopf algebras) and quantum dual (Lie-Poisson) groups~\cite{DrinfeldDuality,SemenovDuality,BallMethod}.
This duality can be expressed in terms of a classical $r$-matrix encoding the coalgebraic properties of a given Hopf algebra in its ``semiclassical'' limit given by the related bialgebra.

Whether or not quantum spacetime models based on these kinds of deformed symmetries arise from more fundamental approaches to quantum gravity remains still an open question.
Some hints that four dimensional quantum gravity could give rise, at low energy limit, to an effective description of quantum spacetime of the kind above mentioned, have been put forward in the literature~\cite{kodadsr,dsrLQGa, GirelliLivineOritikGFT} (see also~\cite{RoncokLQG} for some recent results).
For three dimensional quantum gravity the situation is closer to an assessment.
The fact that in three dimensions gravity~\cite{Witten:1988hc,Achucarro:1987vz} can be described by a topological field theory, which has no dynamical degrees of freedom, renders the theory, and the study of its symmetries, much more manageable.
It follows that three dimensional quantum gravity can be used as a toy model for testing some of the features that are presumed to characterize a (four dimensional) physical theory of quantum gravity.
In such a framework, an effective theory can be achieved both for particles~\cite{Matschull:1997du,Meusburger:2003ta} and for quantum fields~\cite{Freidel:2005me} coupled to gravity.
The study of the symmetries characterizing these effective theories has brought some evidence in favour of a description of an emerging quantum spacetime in terms of $\kappa$-deformed symmetries~\cite{kodadsr, FreidelJurekSmolin, Freidel:2005me} (see also~\cite{TomaszJurekCarroll,TomaszCarroll} for a $\kappa$-deformed Carrollian limit of these theories).

However these preliminary results are still mostly based on semi-heuristic arguments, and most importantly they are missing a complete derivation of the full $\kappa$-Poincar\'e or $\kappa$-de Sitter Hopf algebra (that includes the coalgebraic sector). 
On the other side it has been argued in more systematic analyses relying on a Chern-Simons formulation of 3D gravity coupled to particles that $\kappa$-Poincaré symmetries are indeed not compatible with 3D gravity~\cite{SchrMeuskPoinc}.
Such a conclusion is reached by noticing that the $r$-matrix used in the literature to characterize the $\kappa$-Poincaré (and $\kappa$-de Sitter) Hopf algebras is not compatible with the Ad-invariant bilinear form (the Killing form) associated to the Lie algebra of the gauge group of the Chern-Simons action with cosmological constant, as instead required for the construction of the phase space of particles, described as punctures on the space of Chern-Simons theory~\cite{AlekseevMalkin,Fock:1998nu,SchrMeuskPoinc}.

The aim of this work is to show on the contrary that $\kappa$-de Sitter and $\kappa$-Poincaré symmetries arise naturally in Chern-Simons formulation of (2+1)D gravity with cosmological constant coupled to particles.
The results I obtain, which seem to contradict the previous results~\cite{SchrMeuskPoinc}, are achieved through the definition of a new $r$-matrix which is compatible with the scalar product of Chern-Simons 3D gravity.
The new $r$-matrix proposed in this paper turns out to be the implementation for the Lorentzian case of an $r$-matrix obtained in a recent work by the author where loop quantum gravity (LQG) quantization techniques have been applied to 3D Euclidean gravity~\cite{Cianfrani:2016ogm}.

The manuscript is organized as follows:\\
In Sec.~\ref{sec:deSitter} I define the notation for the symmetry group underlying the Chern-Simons action for (Lorentzian) (2+1)D gravity with cosmological constant: the Lie group generated by the three dimensional de Sitter algebra. 
I show how the group can be recast as SO(3,1), whose Lie algebra can then be splitted in two mutually commuting $\su(2)$ Lie algebras with complex conjugated parameters.\\
In Sec.~\ref{sec:r-matrix} I will briefly outline the Chern-Simons action and construct the new $r$-matrix from the Killing form associated to the Lie algebra of its Gauge (symmetry) group.
\\
The splitting of the group described in Sec.~\ref{sec:deSitter} sets the stage for the following analysis of Sec.~\ref{sec:PoissonLie}, where I will implement the method proposed in~\cite{BallMethod}, which relies on the quantum duality principle, to construct the Hopf algebra $U_q\left( \su_2 \right)$ from the quantization of the dual group associated to the bialgebra of each of the two $\su(2)$ copies.
The Hopf algebra obtained by re-combining together the two $U_q\left( \su_2 \right)$ copies as $U_q\left( \su_2 \right)\oplus U_{q^{-1}}\left( \su_2 \right)$ will describe the deformed symmetry generators for the effective particle theory, the $\kappa$-de Sitter symmetries.\\
In Sec.~\ref{sec:kdSspacetime} I will construct the Hopf algebra dual to $\kappa$-de Sitter, by calculating the Sklyanin Poisson brackets from the invariant vector fields on the two $\su(2)$ copies and their associated $r$-matrix, and then re-combining the two spaces.
The space obtained in this way can be interpreted as the coordinate space for the action of $\kappa$-de Sitter symmetries, i.e. the generalization for $\Lambda \neq 0$ (Lambda being the (positive) cosmological constant) of the three dimensional $\kappa$-Minkowski space.
Finally, in Sec.~\ref{sec:contraction}, I will study the $\Lambda \rightarrow 0$ limit of the $\kappa$-de Sitter Hopf algebra, showing how it contracts to (2+1)D $\kappa$-Poincaré.
It is important to notice that for the coalgebra not to diverge in the $\Lambda \rightarrow 0$ limit it is crucial that the new r-matrix constructed in Sec.~\ref{sec:r-matrix} has the property that it generates two deformed $\su(2)$ copies whose deformation parameters are inverse respect to each other: $q_L=q_R^{-1}=q$.

In the following I will assume units for which the speed of light $c$ as well as the Planck constant $\hbar$ are set to $1$.

\section{Algebra of symmetries in de Sitter spacetime}
\label{sec:deSitter}

In four dimensional gravity, the de Sitter algebra arises as the algebra of spacetime symmetry generators for a solution of Einstein equations describing an homogeneous and isotropic empty universe with positive cosmological constant $\Lambda$.
Indeed in this case the spacetime is maximally symmetric and admits, in four dimensions, 10 symmetry generators that can be identified with the generalization, to a spacetime expanding with a constant rate\footnote{Where $H$ is the Hubble parameter defined in terms of the derivative respect to time of the universe scale $a$.} $H = \dot{a}/a \sim \sqrt{\Lambda}$, of the special relativistic generators of translations, boosts and rotations.
A thorough description of ``de Sitter special relativity'' can be found in~\cite{caccia}, where the construction of the algebra is mostly based on the study of~\cite{Bacry:1968zf}.
In this manuscript I will rely on this physical definition, and define the three dimensional de Sitter group as the reduction to three dimensions of the one reported in~\cite{caccia}.

Denoting time translation, space translations, boosts and rotation respectively as $E$, $P_a$, $N_a$ and $M$, with $a=(1,2)$, the (2+1)D de Sitter group can be described by the Lie algebra~\footnote{Here and in the following, unless otherwise specified, repeated indexes are intended to be summed.}
\begin{equation}
\begin{gathered}\left[E,P_{a}\right]=\Lambda N_{a},\qquad\left[P_{1},P_{2}\right]=\Lambda M,\\
\left[N_{a},E\right]=-P_{a},\qquad\left[N_{a},P_{b}\right]=-\delta_{ab}E,\qquad\left[N_{1},N_{2}\right]=-M,\\
\left[M,N_{a}\right]=\epsilon_{ab}\, N_{b},\qquad\left[M,P_{a}\right]=\epsilon_{ab}\, P_{b},\qquad\left[M,E\right]=0,
\end{gathered}
\label{deSitterAlgebra}
\end{equation}   
where $\Lambda>0$ has dimensions of an inverse length squared.

The de Sitter group in three dimensions is the Lorentz group SO(3,1). By means of the maps
\begin{equation}
E=-\sqrt{\Lambda}K_{3},\quad P_{a}=-\sqrt{\Lambda}\epsilon_{ab}J_{b}, \quad
M=J_{3},\quad N_{a}=-K_{a},
\label{mapPhysLor}
\end{equation} 
the algebra~(\ref{deSitterAlgebra}) is explicitly written as $\so(3,1)$
\begin{equation}
\left[J_{i},J_{j}\right]=\epsilon_{ijk}J_{k},\qquad\left[J_{i},K_{j}\right]=\epsilon_{ijk}K_{k},\qquad\left[K_{i},K_{j}\right]=-\epsilon_{ijk}J_{k},
\label{so(3,1)}
\end{equation} 
$\epsilon_{ijk}$ being the Levi-Civita symbol, with sum over repeated indexes, and $i,j=(1,2,3)$.
A finite dimensional representation can be explicitly obtained in terms of 4x4 real skew-symmetric matrices
\begin{equation}
\left({\cal M}_{AB}\right)_{KL}=\eta_{AK}\delta_{BL}-\eta_{BK}\delta_{AL} \quad \text{where} \quad \eta_{AB}\equiv\text{diag}\{1,-1,-1,-1\} 
\label{so31matrices}
\end{equation} 
with $A,B=(0,1,2,3)$, as
\begin{equation}
\rho\left(J_{i}\right)=\frac{1}{2}\epsilon_{ijk}{\cal M}_{jk},\qquad\rho\left(K_{i}\right)={\cal M}_{0i}.
\label{so31matrixRep}
\end{equation} 
The matrices ${\cal M}_{AB}$ satisfy the commutation rules ($\so(3,1)$)
\begin{equation}
\left[{\cal M}_{AB},{\cal M}_{CD}\right]=\left\{ \eta_{AD}{\cal M}_{BC}+\eta_{BC}{\cal M}_{AD}-\eta_{AC}{\cal M}_{BD}-\eta_{BD}{\cal M}_{AC}\right\} .
\end{equation} 
One can then define an element of SO(3,1) through the exponential map
\footnote{Notice that for $\so(3,1)$ the exponential map is surjective and covers the whole SO(3,1) group.}
\begin{equation}
g=\exp\left(\tfrac{1}{2}\alpha^{AB}{\cal M}_{AB}\right)=\exp\left(j_{i}\rho\left(J_{i}\right)+k_{i}\rho\left(K_{i}\right)\right).
\label{SO31element}
\end{equation} 
Here $j^i$ and $k^i$ are real parameters associated respectively to the ``rotation'' and ``boost'' part of the Lorentz group.
As explicitly shown by the matrices (\ref{so31matrices})-(\ref{so31matrixRep}), in their finite dimensional representation, $J_{i}$ is anti-Hermitian, while $K_{i}$ is Hermitian, accordingly to the fact that the $K_{i}$  sector is non-compact, differently from the $J_{i}$ sector:
\begin{equation}
J_{i}^{*}=-J_{i},\qquad K_{i}^{*}=K_{i}.
\label{realityLorentz}
\end{equation}  

Exploiting the isomorphism\footnote{It can be proved indeed that $\so(3,1)_\C \eq \so(4,C) \eq \su ( 2) \oplus_\C \su ( 2)$} $\so(3,1)_\C \eq \su ( 2) \oplus_\C \su ( 2)$ we complexify the Lie algebra $\so(3,1)$ and choose the basis
\begin{equation}
L_{i}=\frac{1}{2}\left(J_{i}+iK_{i}\right),\qquad R_{i}=\frac{1}{2}\left(J_{i}-iK_{i}\right).
\label{mapLorSU2}
\end{equation} 
These generators satisfy the $\su(2)$ algebra
\begin{equation}
\left[L_{i},L_{j}\right]=\epsilon_{ijk}L_{k},\qquad\left[R_{i},R_{j}\right]=\epsilon_{ijk}R_{k},\qquad\left[L_{i},R_{j}\right]=0. 
\end{equation} 
Thus we have split $\so(3,1)_\C$ in two mutually commuting $\su(2)$ copies, which we call ``left'' and ``right'' copies.
The group element~(\ref{SO31element}) becomes
\begin{equation}
g=g_{l}g_{r}=\exp\left\{ l_{i}\rho\left(L_{i}\right)\right\} \exp\left\{r_{i}\rho\left(R_{i}\right)\right\}
\label{SO31elementSplit}
\end{equation} 
with the two sets of parameters for the left and right copies are related by complex conjugation:
\begin{equation}
l_{i}=j_{i}-ik_{i},\qquad r_{i}=j_{i}+ik_{i}, \qquad l_i = r_i^* .
\label{mapLorSU2param}
\end{equation} 

We further re-express the left and right $\su(2)$ copies in Cartan-Weyl basis as
\begin{equation}
\begin{gathered}
H^{L}=iL_{3},\qquad X_{\pm}^{L}=i\left(L_{1}\pm iL_{2}\right),\\
H^{R}=iR_{3},\qquad X_{\pm}^{R}=i\left(R_{1}\pm iR_{2}\right),
\end{gathered}
\label{mapCartanSU2}
\end{equation} 
closing the algebra
\begin{equation}
\left[H^{I},H^{J}\right]=0,\qquad\left[H^{I},X_{\pm}^{J}\right]=\pm\delta_{IJ}X_{\pm}^{J},\qquad\left[X_{+}^{I},X_{-}^{J}\right]=2\delta_{IJ}H^{J},\qquad I,J=L,R. 
\label{LieAlgebraSU2}
\end{equation} 
From~(\ref{realityLorentz}) we find the reality conditions
\begin{equation}
\tilde{H}=H,\qquad \tilde{X}_{\pm}=X_{\mp}^{*}.
\label{realityCartan}
\end{equation} 
The left and right group elements become respectively
\begin{equation}
\begin{gathered}
g_{l}=\exp\left(\tilde{h}^{L}\rho\left(H^{L}\right)+\tilde{x}_{+}^{L}\rho\left(X_{+}^{L}\right)+\tilde{x}_{-}^{L}\rho\left(X_{-}^{L}\right)\right)\\
g_{r}=\exp\left(\tilde{h}^{R}\rho\left(H^{R}\right)+\tilde{x}_{+}^{R}\rho\left(X_{+}^{R}\right)+\tilde{x}_{-}^{R}\rho\left(X_{-}^{R}\right)\right)
\end{gathered}
\end{equation} 
where
\begin{equation}
\begin{gathered}
\tilde{x}_{+}^{L}=-\frac{i}{2}\left(l^{1}-il^{2}\right),\qquad \tilde{x}_{-}^{L}=-\frac{i}{2}\left(l^{1}+il^{2}\right),\qquad \tilde{h}^{L}=-il^{3},\\
\tilde{x}_{+}^{R}=-\frac{i}{2}\left(r^{1}-ir^{2}\right),\qquad \tilde{x}_{-}^{R}=-\frac{i}{2}\left(r^{1}+ir^{2}\right),\qquad \tilde{h}^{R}=-ir^{3}.
\end{gathered}
\label{mapCartanSU2param}
\end{equation} 
The notation for the coordinate set $\{ \tilde{h},\tilde{x}_{+},\tilde{x}_{-}\}$ will be clarified in the following.

To make contact with the notation used frequently in the 3D gravity literature, we introduce also the generators
\begin{equation}
\begin{gathered}{\cal J}_{0}=-J_{3},\qquad{\cal J}_{a}=K_{a},\\
{\cal P}_{0}=\sqrt{\Lambda}K_{3},\qquad{\cal P}_{a}=\sqrt{\Lambda}J_{a},
\end{gathered}
\label{3Dgenerators}
\end{equation} 
satisfying the algebra
\begin{equation}
\left[{\cal J}_{\mu},{\cal J}_{\nu}\right]=\epsilon_{\mu\nu\rho}{\cal J}^{\rho},\qquad\left[{\cal J}_{\mu},{\cal P}_{\nu}\right]=\epsilon_{\mu\nu\rho}{\cal P}^{\rho},\qquad\left[{\cal P}_{\mu},{\cal P}_{\nu}\right]=-\Lambda\epsilon_{\mu\nu\rho}{\cal J}^{\rho}, 
\label{algebra3Dgen}
\end{equation} 
where $\mu,\nu=(0,1,2)$ and here the indexes have to be raised and lowered through the Lorentzian metric $\eta=\text{diag}(1,-1,-1)$.
The algebra~(\ref{algebra3Dgen}) admits two Casimir
\begin{equation}
C_{1}={\cal P}_{\mu}{\cal P}^{\mu}-\Lambda{\cal J}_{\mu}{\cal J}^{\mu},\qquad C_{2}={\cal P}_{\mu}{\cal J}^{\mu} ,
\label{casimir3D}
\end{equation} 
and satisfy the reality conditions (using~(\ref{algebra3Dgen}) and~(\ref{realityLorentz}))
\begin{equation}
{\cal J}_{0}^{*}=-{\cal J}_{0},\qquad{\cal J}_{a}^{*}={\cal J}_{a},\qquad{\cal P}_{0}^{*}={\cal P}_{0},\qquad{\cal P}_{a}^{*}=-{\cal P}_{a}.
\label{reality3Dgen}
\end{equation}

\section{The Chern-Simons particle action with cosmological constant and a new r-matrix}
\label{sec:r-matrix}

A model of (2+1)D gravity with cosmological constant coupled to particles can be formulated~\cite{Witten:1988hc,AlekseevMalkin,SchrMeuskPoinc} in terms of a Chern-Simons action with de Sitter as gauge group $G$, where, assuming the manifold ${\cal M}$ to be decomposed as the cartesian product $\Sigma\times\mathbb{R}$ of a 2D Riemannian surface (space) and a segment of the real line (time), the particles are described as punctures (conical defects) on $\Sigma$. 
We consider only the case of a single particle and define coordinate $x^0 = t$ on $\mathbb{R}$ and local coordinates $\vec{x}=(x^1,x^2)$ on $\Sigma$, denoting $\vec{x}'$ the puncture coordinates on $\Sigma$.

The gauge field is defined as the Cartan connection on the de Sitter group, i.e. the algebra valued one-form $A \in \g$
\begin{equation}
A=\omega_\mu {\cal J}^\mu + e_\mu {\cal P}^\mu
\end{equation} 
with $\omega_\mu = \omega_\mu^\alpha dx_\alpha$ and $e_\mu = e_\mu^\alpha dx_\alpha$ respectively the spin connection and the dreibein.
The Chern-Simons action for the gauge field $A$ is
\begin{equation}
S_{CS} = \kappa \int_{\cal M} \langle A\wedge dA + \frac{2}{3} A \wedge A \wedge A \rangle_B .
\label{CSaction}
\end{equation} 
Here $\kappa=1/(4\pi G)$ plays the role of the gravity coupling constant, and in (2+1)D has dimensions of a mass. 
The bracket $\langle \cdot \rangle_B$ indicate the inner product~\cite{Witten:1988hc} between all the generators $T_\mu$ of $\g$ in the action respect to the bilinear form corresponding to the second Casimir~(\ref{casimir3D}), explicitly
\begin{equation}
B({\cal J}_\mu,{\cal P}_\nu)=\eta_{\mu\nu}, \quad B({\cal J}_\mu,{\cal J}_\nu)=0, \quad B({\cal P}_\mu,{\cal P}_\nu)=0 ,
\label{AdInvariant}
\end{equation} 
so that for instance the quadratic term in~(\ref{CSaction}) is
\begin{equation}
\int_{\cal M} \langle A\wedge dA \rangle_B = B(T_\mu,T_\nu) \int_{\cal M} \left( A^\mu \wedge dA^\nu \right)
\end{equation} 
Decomposing the connection accordingly to the product structure ${\cal M}=\Sigma\times\mathbb{R}$ as $A=A_0 dt + A_\Sigma$, and introducing the spatial curvature $F_\Sigma = dA_\Sigma + A_\Sigma\wedge A_\Sigma$, the action~(\ref{CSaction}) can be decomposed as
\begin{equation}
S_{CS} = \kappa \int_\mathbb{R} dt \int_{\Sigma} \left\langle \partial_t A_\Sigma \wedge A_\Sigma 
+ A_0 F_\Sigma \right\rangle_B . 
\label{CSaction2+1}
\end{equation} 
The time component of the connection $A_0$ acts as a Lagrange multiplier constraining the curvature to vanish outside the puncture at $\vec{x}'$.

A puncture in ${\cal M}$ is decorated~\cite{AlekseevMalkin,SchrMeuskPoinc} with the action of a free relativistic particle.
The particle's degrees of freedom are encoded in an element $\xi_0$ of the Lie algebra $\mathfrak{g}^*$ dual to the Lie algebra $\mathfrak{g}$ of $G$, defined by the coadjoint orbits of $G$. 
Explicitly, in terms of the generators dual to~(\ref{3Dgenerators}),
\begin{equation}
\xi_0 = m \tilde{{\cal P}}_0 + s \tilde{{\cal J}}_0 \ , 
\qquad \xi = \text{Ad}^* \xi_0 = p^\mu \tilde{{\cal P}}_\mu + j^\mu \tilde{{\cal J}}_\mu \ :
\label{coadjointOrbit}
\end{equation}
$\xi_0$ fixes the orbit by giving the values of the rest mass $m$ and spin $s$ of the particle, while $\xi$,  obtained through the coadjoint action of $G$ on $\xi_0$, encodes a generic state of motion characterized by momentum $p^\mu$ and angular momentum $j^\mu$.
The generators $\tilde{{\cal P}}_\mu, \tilde{{\cal J}}_\mu$ form a basis $\{\tilde{e}_i\}$ of $\g^*$, satisfying the canonical duality relations $\langle \tilde{e}_i, e_j\rangle=\delta_{ij}$ with the basis set ${\cal P}_\mu, {\cal J}_\mu = \{e_i\}$ of $\g$.
The coadjoint action of an element $g\in G$ is defined by the relation $\langle \Ad^*_g \tilde{Y},X \rangle = \langle \tilde{Y} , g^{-1} X g\rangle$ for $X\in\g$, $\tilde{Y}\in\g^*$.
The dual generators are mapped to $\g$ by the map $\phi:\g^*\rightarrow\g$, which must be compatible with the $\Ad$-invariant bilinear form on $\g$, $B:\g\times\g\rightarrow\C$, so that $\langle \tilde{Y} , X \rangle = B(\phi(\tilde{Y}),X)$.
From~(\ref{AdInvariant}) it follows
\begin{equation}
\phi(\tilde{{\cal J}}_\mu)= P_\mu,\quad  \phi(\tilde{{\cal P}}_\mu)= J_\mu \ ,
\end{equation} 
and the free particle action is
\begin{equation}
\int_\mathbb{R} dt \left\langle \phi(\xi_0) g^{-1} \partial_t g \right\rangle_B .
\end{equation} 
The particle action is minimally coupled to the Gauge field~\cite{AlekseevMalkin,SchrMeuskPoinc} so that the total action is
\begin{equation}
\begin{split}
S = \int_\mathbb{R} dt {\cal L}, \qquad  
{\cal L} = & \kappa \int_\Sigma \left\langle \partial_t A_\Sigma \wedge  A_\Sigma \right\rangle_B 
- \left\langle \phi(\xi_0) g^{-1} \partial_t g \right\rangle_B \\
& + \int_{\Sigma} \left\langle A_0 \left( \kappa F_\Sigma
- \phi(\xi) \delta^2(\vec{x}-\vec{x}') dx^1 \wedge dx^2 \right) \right\rangle_B .
\end{split}
\label{CSaction}
\end{equation} 
It can be shown (see for instance~\cite{Bimonte:1997dw} for an explicit derivation) that this action reduces in its metric formulation to the action of (2+1)D gravity with cosmological constant coupled to a point particle.

In this construction the particle phase space variables corresponding to momenta\footnote{I refer in general to particle momenta denoting the whole set of energy, spatial momentum, angular momentum and boost charges corresponding to translations and Lorentz transformations.} are described~\cite{AlekseevMalkin} by elements of the Poisson-Lie group $G^*$ associated to the coadjoint orbit~(\ref{coadjointOrbit}).
The deformation quantization of the algebra of momenta can be obtained 
(see ~\cite{SemenovDuality}, Chapter 8 of~\cite{MajidFound}, and also~\cite{SchrMeuskPoinc,BallMethod}) by relating the Poisson-Lie group to the corresponding (coboundary) Lie bialgebra $(\g,\delta)$, where $\delta$ is the co-commutator obtained from the $r$-matrix $r$ associated to $\g$ as\footnote{The notation is $r=\sum_i r^{(1)}_i\ot r^{(2)}_i = \sum r^{(1)}\ot r^{(2)}$ with the summation indexes omitted.}
\begin{equation}
\delta(X) = \sum ([X,r_{(1)} ]\otimes 1  + 1\otimes [X,r_{(2)}]) .
\label{cocommutator}
\end{equation} 
The $r$-matrix must be compatible with the bilinear invariant form~(\ref{AdInvariant}) so that its symmetrical part $r_+$ is proportional to its  (tensorized) Casimir $r_+ \propto \frac{1}{2}B^{ij}e_i\otimes e_j$.
Fixing the antisymmetrical part $r_-$ of $r$ so that the classical Yang-Baxter equation (CYBE) is satisfied, the associated Lie bialgebra is coboundary and quasi-triangular, and is the semi-classical limit of a quantum group of symmetries in the sense of Hopf-algebras~(see e.g.~\cite{MajidFound}, Ch.~8.1).

In the basis $({\cal J}_\mu,{\cal P}_\mu)$ the Casimir is $C_2$ of~(\ref{casimir3D}) and the symmetric part of the r-matrix must be proportional to (remember that indexes are raised and lowered with the metric $\eta_{\mu\nu} = \text{diag}(1,-1,-1)$ and repeated indexes are summed)
\begin{equation}
r_+ = {\cal J}_{\mu}\otimes{\cal P}^{\mu}+{\cal P}_{\mu}\otimes{\cal J}^{\mu} \ .
\end{equation} 
The CYBE\footnote{The notation is such that $r_{ij} = \sum1\ot\cdots\ot r^{(1)}\ot1\ot\cdots\ot r^{(2)}\ot\cdots\ot1$ is $r$ in its $i$th and $j$th factor.} $[\![ r,r]\!] = [r_{12},r_{13}]+[r_{12},r_{23}]+[r_{13},r_{23}]=0$ is satisfyied if the antisymmetric part of the r-matrix is
\begin{equation}
r_- = i m_{\rho}\epsilon^{\rho\mu\nu}\left({\cal J}_{\mu}\otimes{\cal P}_{\nu}+{\cal P}_{\mu}\otimes{\cal J}_{\nu}\right) \ ,
\end{equation} 
with $m_\mu$ a unit timelike vector $m^\mu m_\mu =1$, that can be fixed to $m_\mu = (1,0,0)$.

The $r$-matrix is deformation-quantized by introducing a quantum deformation parameter (see e.g.~\cite{BallMeuskdS}) $\zz$ so that the $r$-matrix becomes $r_\zz = \zz(r_+ + r_-)$.
The explicit form of the deformation parameter is determined by the following requirements:
\begin{enumerate}
 \item 
Since we are working with real Lie bialgebras, we want the anti-symmetric part of the $r$-matrix to be real, so that the co-commutators are also real and they generate a real dual Lie algebra (see next section).
It follows that $\zz$ must be purely imaginary.
\item  The dimension of the deformation parameter is determined~\cite{Ball19943D,BallMeuskdS} by the ``primitive'', in the sense of non deformed, time-translation generator ${\cal P}_0$, so that $[\zz] = [{\cal P}_0]^-1$, as it will become clearer in the following sections. Thus $\zz$ must have dimensions of a mass, i.e. it must be proportional to $\kappa$.
This corresponds with the definition of a ``time-like'' r-matrix in the language of~\cite{Ball19943D,BallMeuskdS}.
\end{enumerate}
Substituting then $\zz=i/\kappa$ the deformed r-matrix is
\begin{equation}
r_\zz = \frac{i}{\kappa} \left( {\cal J}_{\mu}\otimes{\cal P}^{\mu}+{\cal P}_{\mu}\otimes{\cal J}^{\mu} \right) 
- \frac{1}{\kappa} \epsilon^{ij}\left({\cal J}_{i}\otimes{\cal P}_{j}+{\cal P}_{i}\otimes{\cal J}_{j}\right) \ .
\label{rmatrix}
\end{equation} 
Notice that the hermiticity of the generators~(\ref{reality3Dgen}) are such that the r-matrix satisfy a well-defined reality condition $r^{(*\ot*)}=\tau(r)$, where $\tau$ is the flip operator ($\tau(a\ot b) = (b\ot a)$). This means~\cite{MajidFound} that the corresponding quantum ${\cal R}$-matrix is ``real'', and the associated ``quantum inverse Killing form'' ${\cal Q}={\cal R}_{21}{\cal R}$ is self-adjoint.

We can now use the splitting of the algebra defined in the previous section to rewrite the $r$-matrix in terms of the two copies of $\su_2$ generators. Combining~(\ref{3Dgenerators}) with~(\ref{mapLorSU2}) and~(\ref{mapCartanSU2}) we obtain
\begin{equation}
r_\zz=\frac{2\sqrt{\Lambda}}{\kappa}\left(H^{L}\otimes H^{L}-H^{R}\otimes H^{R}+X_{+}^{L}\otimes X_{-}^{L}-X_{+}^{R}\otimes X_{-}^{R}\right) 
\label{rMatrixCartan}
\end{equation} 
Due to the fact that the two $\su(2)$ copies are mutually commuting, this $r$-matrix is the sum of the contributions of the $r$-matrices for the two $\su(2)$ copies: $r_z=r_{L}+r_{R}$ where $\R=\R_{l}\R_{r}$ , and we can consider them separately.
Notice that for each of the two copies it has the form
\begin{equation}
r_{L,R}= \z_{L,R} \left(H^{L,R}\otimes H^{L,R}+X_{+}^{L,R}\otimes X_{-}^{L}\right),
\qquad \z_L = - \z_R = 2\sqrt{\Lambda}/\kappa .
\label{rMatrixSU2}
\end{equation} 
From~(\ref{realityCartan}) it follows that $r_{L,R}^{(*\ot *)} = \tau(r)$ and $\z$ is real.
These reality conditions coincide with the ones for $U_q(\su(2))$, where $q=\exp{\z}$ is real and $\R^{(*\ot *)}=\tau(\R)$~(see \cite{MajidFound}, Ch~3).
We will see indeed in the next section that we will recover these Hopf algebras upon quantization.

The $r$-matrix (\ref{rMatrixCartan}) turns out to be the Lorentzian version of the $r$-matrix obtained in~\cite{Cianfrani:2016ogm} for the 3D Euclidean case through LQG quantization techniques.
In~\cite{Cianfrani:2016ogm} it was underlined how the opposite sign of the deformation parameters of the two quantized $\su$(2) copies, which in that context arose from the the quantization of the holonomy relative to each of the two copies, is necessary for the convergence of the contraction limit $\Lambda\rightarrow 0$ of the Hopf algebra of symmetries.
We will obtain an analogous result for the Lorentzian case in Sec.~\ref{sec:contraction}.

\section{Derivation of the $\kappa$-de Sitter algebra of symmetries}
\label{sec:PoissonLie}

As stated above, in the approach outlined in the previous section, once the coadjoint orbit of $G$ is fixed by the values of the particle's mass and spin, the particle's momenta (and angular momenta) are the parameters of the dual Poisson-Lie group $G^*$.
Its infinitesimal counterpart is the Lie bialgebra $(\g^*,\delta^*)$ obtained from the deformed r-matrix $r_z$.
As explained in detail in~\cite{BallMethod}, by virtue of the quantum-duality principle, which establishes a correspondence between a quantum universal enveloping algebra (a Hopf algebra) and a quantum dual group, the quantization as a Hopf-algebra of $G^*$ together with its Poisson structure provides the Hopf algebra $(U_z(\g),\Delta_z)$. 
Finally, this will be the (Hopf) algebra of the symmetry generators corresponding to momenta and angular momenta of the particle, i.e. the time and space translation and Lorentz generators.
The aim of this section is then to evaluate this (Hopf) algebra relying on the splitting of the de Sitter group in two $\su(2)$ copies described in the previous sections and in the method introduced in~\cite{BallMethod}.

\subsection{The dual Lie bialgebra $\g^*$ and the dual group $G^*$}

Starting from the $r$-matrix~(\ref{rMatrixSU2}) one can construct the (coboundary) Lie bialgebra $(\g_{L,R},\delta_{L,R})$ for each of the two $\su(2)$ copies introduced in the previous sections. 
Since, apart from the sign of the deformation parameter, the $r$-matrix has the same form for each of the two copies, in the following of this section I will omit the subscript $L$ or $R$ denoting the structures related to the two $\su(2)$ copies, which however must not be confused with the structures related to the de Sitter algebra $\g$ and group $G$ used in the other sections.

A Lie bialgebra is defined~(see~\cite{MajidFound}, Ch.~8) by the Lie algebra $\mathfrak{g}$ and the cocommutators $\delta$ through the structure constants $c_{ij}^{\ \ k}$
and $f_{\ \ k}^{ij}$ as
\begin{equation}
\begin{gathered}\mathfrak{g}:\quad\left[e_{i},e_{j}\right]=c_{ij}^{\ \ k}e_{k},\qquad\delta:\quad\delta\left(e_{i}\right)=\end{gathered}
f_{\ \ k}^{ij}e_{i}\ot e_{j}.
\end{equation}
For the case under consideration the structure constants $c_{ij}^{\ \ k}$ are given by~(\ref{LieAlgebraSU2}), while from~(\ref{LieAlgebraSU2}) and~(\ref{rMatrixSU2}) we find, using~(\ref{cocommutator}),
\begin{equation}
\delta\left(H\right)=0,\qquad\delta\left(X_{+}\right)=zX_{+}\wedge H,\qquad\delta\left(X_{-}\right)=zX_{-}\wedge H ,
\end{equation}
where $a\wedge b=a\otimes b-b\otimes a$.
The fact that the bialgebra is coboundary is guaranteed by the $\Ad$-invariance of the symmetric part of its $r$-matrix.

The dual Lie bialgebra~(see~\cite{MajidFound} Ch.~8) $\left(\mathfrak{g}^{*},\delta^{*}\right)$ with
basis $\left\{ e^{i}\right\} =\left\{ \tilde{e}_{i}\right\} $ is defined by dualisation according to
\begin{equation}
\left\langle [ \tilde{e}_i,\tilde{e}_j ], e_k \right\rangle 
=  \left\langle \tilde{e}_i \ot \tilde{e}_j , \delta({e}_k) \right\rangle , \qquad 
\left\langle \delta({\tilde{e}}_k) , e_i \ot e_j \right\rangle 
= \left\langle \tilde{e}_k , [ e_i,e_j ]\right\rangle .
\end{equation} 
This amounts, considering the canonical dualization $< \tilde{e}_i, e_j> = \delta_{ij}$, to interchange the role of the structure constants:
\begin{equation}
\mathfrak{g}^{*}:\quad\left[e^{i},e^{j}\right]=f_{\ \ k}^{ij}e^{k},\qquad\delta^{*}:\quad\delta^{*}\left(e^{i}\right)=c_{jk}^{\ \ i}e^{j}\ot e^{k}.
\end{equation}
In the basis $\left\{ e^{i}\right\} :\left\{ \tilde{H},\tilde{X}_{+},\tilde{X}_{-}\right\} $
we thus find
\begin{equation}
\left[\tilde{H},\tilde{X}_{\pm}\right]=-z\tilde{X}_{\pm},\qquad\left[\tilde{X}_{+},\tilde{X}_{-}\right]=0,
\label{LieAlgebraSU2dual}
\end{equation}
\begin{equation}
\delta^{*}\left(\tilde{H}\right)=2\tilde{X}_{+}\wedge \tilde{X}_{-},\qquad\delta^{*}\left(\tilde{X}_{\pm}\right)=\pm \tilde{H}\wedge \tilde{X}_{\pm}.\label{eq:dualCocommutators}
\end{equation}

Following the line of reasoning of~\cite{BallMethod}, in order to know if $\left(\mathfrak{g}^{*},\delta6\right)$ is coboundary, we must check if there exists an r-matrix $\tilde{r}$ whose skew-symmetric part generates the cocommutators (\ref{eq:dualCocommutators})
through equation (\ref{cocommutator}). It is easy to show
by parametrizing
\begin{equation}
\tilde{r}_{A}=\alpha \tilde{H}\wedge \tilde{X}_{+}+\beta \tilde{H}\wedge \tilde{X}_{-}+\gamma \tilde{X}_{+}\wedge \tilde{X}_{-}
\end{equation}
that the equation
\begin{equation}
\delta^{*}\left(e^{i}\right)=\tilde{r}_{A(1)}\otimes\left[e^{i},\tilde{r}_{A(2)}\right]+\left[e^{i},\tilde{r}_{A(1)}\right]\otimes \tilde{r}_{A(2)}
\end{equation}
with $\delta^{*}\left(e^{i}\right)$ given by (\ref{eq:dualCocommutators})
has no solution. Thus the bialgebra $\left(\mathfrak{g}^{*},\delta^{*}\right)$
is non-coboundary.
This implies that, in order to evaluate the Poisson structure on $G^*$, one cannot use the definition of the Sklyanin bracket, which, for an $r$-matrix $r=r^{ij}e_i\ot e_j$ in a basis $\{e_i\}$ of a Lie algebra $\g$ is
\begin{equation}
\left\lbrace a, b \right\rbrace = \frac{1}{2} r^{ij} \cdot((X_i^R\wedge X^R_j - X_i^L\wedge X^L) (a\ot b)),
\label{Sklyanin}
\end{equation} 
where $X^R_i$ and $X^L_i$ are respectively right and left vector fields for the generator $e_i$.

In~\cite{BallMethod} an alternative method was proposed to evaluate the Poisson structure on $G^*$. For the case under consideration the problem can be solved algebraically.
Starting from the adjoint representation of $\mathfrak{g}^{*}$ $\left(\rho\left(\tilde{e}_{i}\right)\right)_{k}^{j}=-f_{\ \ k}^{ij}$,
from which we get
\begin{equation}
\rho\left(\tilde{H}\right)=\left(\begin{array}{ccc}
0 & 0 & 0\\
0 & z & 0\\
0 & 0 & z
\end{array}\right),\qquad\rho\left(\tilde{X}_{+}\right)=\left(\begin{array}{ccc}
0 & -z & 0\\
0 & 0 & 0\\
0 & 0 & 0
\end{array}\right),\qquad\rho\left(\tilde{X}_{-}\right)=\left(\begin{array}{ccc}
0 & 0 & -z\\
0 & 0 & 0\\
0 & 0 & 0
\end{array}\right),
\end{equation}
a generic group element of $G^{*}$ can be constructed
as the ordered product of exponentials
\begin{equation}
g^{*}=\exp\left(h\rho\left(\tilde{H}\right)\right)\exp\left(x_{+}\rho\left(\tilde{X}_{+}\right)\right)\exp\left(x_{-}\rho\left(\tilde{X}_{-}\right)\right)=\left(\begin{array}{ccc}
1 & -zx_{+} & -zx_{-}\\
0 & e^{zh} & 0\\
0 & 0 & e^{zh}
\end{array}\right).
\end{equation}
Notice that we could consider a different ordering prescription, corresponding
to a different parametrization of the group, as for instance
\begin{equation}
\begin{gathered}
g^{*} =  \exp\left(\tilde{h}\rho\left(\tilde{H}\right)/2\right)\exp\left(\tilde{x}_{+}\rho\left(\tilde{X}_{+}\right)\right)\exp\left(\tilde{x}_{-}\rho\left(\tilde{X}_{-}\right)\right)\exp\left(\tilde{h}\rho\left(\tilde{H}\right)/2\right) \\
= \left(\begin{array}{ccc}
1 & -ze^{z\tilde{h}/2}\tilde{x}_{+} & -ze^{z\tilde{h}/2}\tilde{x}_{-}\\
0 & e^{z\tilde{h}} & 0\\
0 & 0 & e^{z\tilde{h}}
\end{array}\right).
\end{gathered}
\label{AlternativeParam}
\end{equation}
The two sets of coordinates are connected by the map
\begin{equation}
\tilde{h}=h,\qquad\tilde{x}_{\pm}=e^{-zh/2}x_{\pm}.\label{eq:tildeCoord}
\end{equation}

In order to define the Hopf-algebra on $G^*$ we must construct the coproduct $\Delta_{G^{*}}$ (I will omit in the following the subscript $G^{*}$). 
This can be derived solving a set of functional equations which reflect the fact that the coproduct map for coordinate functions on $G^{*}$ is the pullback of the group multiplication law in (dual) algebraic form, so that the coassociativity of the coproduct is the associativity of group multiplication (see~\cite{BallMethod} for the details).
In our case this amounts to solve the equation
\begin{equation}
\begin{gathered}
\left(\begin{array}{ccc}
\Delta1 & -z\Delta x_{+} & -z\Delta x_{-}\\
0 & \Delta e^{zh} & 0\\
0 & 0 & \Delta e^{zh}
\end{array}\right)= ~~~~~~~~~~~~~~~~~~~~~~~~~~~~~ ~~~~~~~~~~~~~~~~~~~~~~~~~~~~~~~~~~~~~~~~~~~\\
\left(\begin{array}{ccc}
1\otimes1 & -zx_{+}\otimes1 & -zx_{-}\otimes1\\
0 & e^{zh}\otimes1 & 0\\
0 & 0 & e^{zh}\otimes1
\end{array}\right) \left(\begin{array}{ccc}
1\otimes1 & -z1\otimes x_{+} & -z1\otimes x_{-}\\
0 & 1\otimes e^{zh} & 0\\
0 & 0 & 1\otimes e^{zh}
\end{array}\right)
\end{gathered}
\end{equation}
from which we get
\begin{equation}
\Delta1=1\otimes1,\qquad\Delta h=h\otimes1+1\otimes h,\qquad\Delta x_{\pm}=x_{\pm}\otimes e^{zh}+1\otimes x_{\pm}.\label{eq:coproducts}
\end{equation}
Notice that in the alternative parametrization of $G^{*}$ defined in~(\ref{AlternativeParam}) we would have
\begin{equation}
\Delta1=1\otimes1,\qquad\Delta\tilde{h}=\tilde{h}\otimes1+1\otimes\tilde{h},\qquad\Delta\tilde{x}_{\pm}=\tilde{x}_{\pm}\otimes e^{z\tilde{h}/2}+e^{-z\tilde{h}/2}\otimes\tilde{x}_{\pm}.
\end{equation}

\subsection{The Poisson structure on $G^{*}$}

The Poisson structure $\lambda$ on $G^{*}$ has to satisfy two requirements:
\begin{itemize}
\item The group (co)multiplication has to be a Poisson map respect to $\lambda$
\begin{equation}
\left\{ \Delta_{G^{*}}\left(a\right),\Delta_{G^{*}}\left(b\right)\right\} _{\lambda}=\Delta_{G^{*}}\left(\left\{ a,b\right\} _{\lambda}\right)\qquad\forall a,b\in G^{*}\label{eq:homomorphism}
\end{equation}
\item The linearization of $\lambda$ should coincide with the Lie algebra
defined by the structure tensor $c_{ij}^{\ \ k}$ defining $\delta^{*}$.
\end{itemize}

First we assume that the brackets are of the form
\begin{equation}
\left\{ x_{i},x_{j}\right\} _{\lambda}=Q_{ij}=\sum_{k,l}\beta_{ijkl}F_{k}F_{l}
\end{equation}
where $\beta_{ijkl}$ are arbitrary coefficients and $F_{i}$
are among the set of functions appearing as matrix entries of group
elements of $G^{*}$ and in the coproducts for the coordinates $x_{i}$: 
\begin{equation}
\left\{ F_{i}\right\} :=\left\{ 1,h,x_{+},x_{-},e^{zh}\right\} .
\end{equation}
I.e. the Poisson brackets are homogeneous quadratic in terms of functions
included within the set $\left\{ F_{i}\right\} $. 
The homomorphism condition (\ref{eq:homomorphism}) becomes
\begin{equation}
\sum_{k,l=1}^{d}\left(\left(Q_{kl}\otimes1\right)\left(\frac{\partial\Delta x_{i}}{\partial\left(x_{k}\otimes1\right)}\cdot\frac{\partial\Delta x_{j}}{\partial\left(x_{l}\otimes1\right)}\right)+\left(1\otimes Q_{kl}\right)\left(\frac{\partial\Delta x_{i}}{\partial\left(1\otimes x_{k}\right)}\cdot\frac{\partial\Delta x_{j}}{\partial\left(1\otimes x_{l}\right)}\right)\right)=\Delta Q_{ij}.
\end{equation}
Let's consider the equation term by term. For the terms $Q_{0+}$
and $Q_{0-}$ we have 
\begin{equation}
\begin{gathered}\Delta Q_{0+}=Q_{0+}\otimes e^{zh}+1\otimes Q_{0+},\\
\Delta Q_{0-}=Q_{0-}\otimes e^{zh}+1\otimes Q_{0-}
\end{gathered}
\end{equation}
These are easily solved by
\begin{equation}
\begin{gathered}Q_{0+}=\alpha_{0+}\left(e^{zh}-1\right)+\beta_{0+}x_{+}+\gamma_{0+}x_{-},\\
Q_{0-}=\alpha_{0-}\left(e^{zh}-1\right)+\beta_{0-}x_{+}+\gamma_{0-}x_{-},
\end{gathered}
\label{eq:PLbrackets1}
\end{equation}
where I renamed the constant coefficients to be determined. The term
$Q_{+-}$ gives 
\begin{equation}
\Delta Q_{+-}=Q_{+-}\otimes e^{2zh}+1\otimes Q_{+-}+zx_{+}\otimes Q_{0-}e^{zh}-zx_{-}\otimes Q_{0+}e^{zh}.
\end{equation}
This is solved by
\begin{equation}
Q_{+-}=\alpha_{+-}\left(e^{2zh}-1\right)-z\alpha_{0+}x_{-}+z\alpha_{0-}x_{+}+\frac{z}{2}\beta_{0-}x_{+}^{2}-\frac{z}{2}\gamma_{0+}x_{-}^{2}-z\beta_{0+}x_{+}x_{-}\label{eq:PLbrackets2}
\end{equation}
and
\begin{equation}
\gamma_{0-}=-\beta_{0+}.
\end{equation}
Thus we are left with 6 parameters to be determined. We must impose
now the second condition, the one on the linearization of the brackets.
I.e. we must impose that
\begin{equation}
\sum_{k=1}^{d}\frac{\partial Q_{ij}}{\partial x_{k}}\Big|_{\left\{ x_{i}\right\} =0}x_{k}=c_{ij}^{\ \ k}x_{k}.
\end{equation}
At linear order, the brackets (\ref{eq:PLbrackets1}) and (\ref{eq:PLbrackets2})
become
\begin{equation}
\begin{gathered}\left\{ h,x_{+}\right\} _{0}=z\alpha_{0+}h+\beta_{0+}x_{+}+\gamma_{0+}x_{-},\\
\left\{ h,x_{-}\right\} _{0}=z\alpha_{0-}h+\beta_{0-}x_{+}-\beta_{0+}x_{-},\\
\left\{ x_{+},x_{-}\right\} _{0}=2z\alpha_{+-}h-z\alpha_{0+}x_{-}+z\alpha_{0-}x_{+},
\end{gathered}
\end{equation}
Comparing the brackets with (\ref{LieAlgebraSU2}) we find
\begin{equation}
\begin{gathered}\alpha_{0+}=\gamma_{0+}=\alpha_{0-}=\beta_{0-}=0,\\
\beta_{0+}=1,\qquad\alpha_{+-}=\frac{1}{z}.
\end{gathered}
\end{equation}
so that we finally obtain
\begin{equation}
\left\{ h,x_{\pm}\right\} =\pm x_{\pm},\qquad\left\{ x_{+},x_{-}\right\} =\frac{e^{2zh}-1}{z}-zx_{+}x_{-}.
\end{equation}

Together with the coproducts (\ref{eq:coproducts}), these brackets
define the Poisson-Hopf algebra associated to the dual Lie bialgebra
$(\mathfrak{g}^{*},\delta^{*})$. In terms of the alternative set
of coordinates $\tilde{x}_{i}$ defined in (\ref{eq:tildeCoord})
the Poisson-Lie brackets are 
\begin{equation}
\left\{ \tilde{h},\tilde{x}_{\pm}\right\} =\pm\tilde{x}_{\pm}\qquad\left\{ \tilde{x}_{+},\tilde{x}_{-}\right\} =2\frac{\sinh\left(z\tilde{h}\right)}{z}.
\end{equation}
Even though the calculations were easier in the basis $\left\{ x_{i}\right\} $,
the basis $\left\{ \tilde{x}_{i}\right\} $, due to its symmetries,
is more suitable for quantization, as I will show in the next section.

\subsection{Quantization as a Hopf-algebra}

The quantization as a Hopf-algebra of the Poisson-Hopf group $(G^*,\lambda,\Delta)$ consists in promoting the group coordinates to non-commutative generators. 
In general we have to face ordering ambiguities. 
However, contrary to the basis $\left\{ x_{i}\right\} $, where on the r.h.s. of the Poisson-Lie brackets there appear products of $x_{+}$ and $x_{-}$, in the basis of coordinates $\left\{ \tilde{x}_{i}\right\} $ the non-linear terms on the r.h.s. are all functions of coordinates which have vanishing Poisson brackets between themselves. It thus follows that in the basis $\left\{ \tilde{x}_{i}\right\} $ the quantization is straightforwardly obtained by substituting the Poisson brackets $\left\{\cdot , \cdot \right\} $ with the commutators $ \left[\cdot , \cdot\right]$ between the Hopf-algebra generators.
I will choose this basis and indicate the Hopf generators with the same symbols used for the ones of the starting Lie algebra $(H,X_+,X_-)$, since they will describe a combination of the generators of the deformed relativistic symmetries for the particle, in the same way the $\su(2)$ generators were a combination of the symmetries of de Sitter spacetime (see Sec.~\ref{sec:deSitter}).

Finally, we have obtained the Hopf algebra of generators
\begin{equation}
\begin{gathered}\begin{gathered}\left[H,X_{\pm}\right]=\pm X_{\pm},\qquad\left[X_{+},X_{-}\right]=\frac{\sinh\left(zH\right)}{z/2},\\
\Delta H=H\otimes1+1\otimes H,\qquad\Delta X_{\pm}=X_{\pm}\otimes e^{zH/2}+e^{-zH/2}\otimes X_{\pm}
\end{gathered}
\end{gathered}
\end{equation}
Notice that if we rescale the generators as
\begin{equation}
X_{\pm}\rightarrow\sqrt{\frac{z/2}{\sinh\left(z/2\right)}}X_{\pm}
\end{equation}
we obtain the $SU_{q}\left(2\right)$ algebra in its usual form (as
in Majid's book)
\begin{equation}
\begin{gathered}\left[H,X_{\pm}\right]=\pm X_{\pm},\qquad\left[X_{+},X_{-}\right]=\frac{\sinh\left(zH\right)}{\sinh\left(z/2\right)},\\
\Delta H=H\otimes1+1\otimes H,\qquad\Delta X_{\pm}=X_{\pm}\otimes e^{zH/2}+e^{-zH/2}\otimes X_{\pm} .
\end{gathered}
\label{HopfSU2}
\end{equation}

We can now obtain the Hopf algebra of the generators of time and space translations, boosts and rotations, combining the maps~(\ref{mapPhysLor}), (\ref{mapLorSU2}) and~(\ref{mapCartanSU2}) as
\begin{equation}
\begin{gathered}
E=\sqrt{\Lambda}\left(H^{L}-H^{R}\right),\qquad M=-i\left(H^{L}+H^{R}\right),\\
P_{1}=\frac{\sqrt{\Lambda}}{2}\left(X_{+}^{L}-X_{-}^{L}+X_{+}^{R}-X_{-}^{R}\right),\\
P_{2}=-\frac{i\sqrt{\Lambda}}{2}\left(X_{+}^{L}+X_{-}^{L}+X_{+}^{R}+X_{-}^{R}\right),\\
N_{1}=\frac{1}{2}\left(X_{+}^{L}+X_{-}^{L}-X_{+}^{R}-X_{-}^{R}\right),\\
N_{2}=-\frac{i}{2}\left(X_{+}^{L}-X_{-}^{L}-X_{+}^{R}+X_{-}^{R}\right).
\end{gathered}
\label{mapPhysCartan}
\end{equation} 
Considering that relations~(\ref{HopfSU2}) hold for both $L$ and $R$ copies, but keeping in mind that $z_L=-z_R=2\sqrt{\Lambda}/\kappa$, we obtain, combining~(\ref{mapPhysCartan}) with~(\ref{HopfSU2}) for the respective $L$ and $R$ $U_q(\su(2))$ copies,
\begin{equation}
\begin{gathered}\left[E,P_{a}\right]=\Lambda N_{a},\qquad\left[N_{a},E\right]=-P_{a},\\
\left[P_{1},P_{2}\right]=\Lambda\frac{\sin\left(\sqrt{\Lambda}M/\kappa\right)}{\sinh\left(\sqrt{\Lambda}/\kappa\right)}\cosh\left(E/\kappa\right),\\
\left[N_{a},P_{b}\right]=-\delta_{ab}\sqrt{\Lambda}\frac{\sinh\left(E/\kappa\right)}{\sinh\left(\sqrt{\Lambda}/\kappa\right)}\cos\left(\sqrt{\Lambda}M/\kappa\right),\\
\left[N_{1},N_{2}\right]=-\frac{\sin\left(\sqrt{\Lambda}M/\kappa\right)}{\sinh\left(\sqrt{\Lambda}/\kappa\right)}\cosh\left(E/\kappa\right),\\
\left[M,N_{a}\right]=\epsilon_{a}^{\ b}N_{b},\qquad\left[M,P_{a}\right]=\epsilon_{a}^{\ b}P_{b},\qquad\left[M,E\right]=0.
\end{gathered}
\label{kdeSitterAlg}
\end{equation} 
\begin{equation}
\begin{gathered}\Delta E=E\otimes1+1\otimes E\ ,\qquad\Delta M=M\otimes1+1\otimes M\ ,\\
\begin{split}\Delta P_{a}= & P_{a}\otimes e^{\frac{1}{2}E/\kappa}\cos\left(\frac{\sqrt{\Lambda}}{2\kappa}M\right)+e^{-\frac{1}{2}E/\kappa}\cos\left(\frac{\sqrt{\Lambda}}{2\kappa}M\right)\otimes P_{a}\\
 & -\epsilon_{ab}\,\left(\sqrt{\Lambda}N_{b}\otimes e^{\frac{1}{2}E/\kappa}\sin\left(\frac{\sqrt{\Lambda}}{2\kappa}M\right)-\sqrt{\Lambda}e^{-\frac{1}{2}E/\kappa}\sin\left(\frac{\sqrt{\Lambda}}{2\kappa}M\right)\otimes N_{b}\right)\,,
\end{split}
\\
\begin{split}\Delta N_{a}= & N_{a}\otimes e^{\frac{1}{2}E/\kappa}\cos\left(\frac{\sqrt{\Lambda}}{2\kappa}M\right)+e^{-\frac{1}{2}E/\kappa}\cos\left(\frac{\sqrt{\Lambda}}{2\kappa}M\right)\otimes N_{a}\\
 & -\epsilon_{ab}\,\left(\frac{1}{\sqrt{\Lambda}}P_{b}\otimes e^{\frac{1}{2}E/\kappa}\sin\left(\frac{\sqrt{\Lambda}}{2\kappa}M\right)-\frac{1}{\sqrt{\Lambda}}e^{-\frac{1}{2}E/\kappa}\sin\left(\frac{\sqrt{\Lambda}}{2\kappa}M\right)\otimes P_{b}\right)\,.
\end{split}
\end{gathered}
\label{kdeSitterCopr}
\end{equation} 
These relations define the Hopf algebra, which I denote $\kappa$-de Sitter, characterizing the relativistic symmetry generators of a particle in (2+1)D (Lorentzian) gravity with cosmological constant.

\section{The non-commutative $\kappa$-deSitter spacetime}
\label{sec:kdSspacetime}

In the previous sections we derived the deformed algebra of symmetry generators, which we denoted as $\kappa$-de Sitter, by ``quantizing'' the Poisson brackets between the coordinates of the dual group $G^*_{L,R}$ for each of the $L$ and $R$ copies.
Since the Lie bialgebra $(\g^*,\delta^*)$ is not coboundary, we followed an analytic procedure to derive the Poisson structure on $G^*_{L,R}$.
For the coordinates of the group $G_{L,R}$, whose Lie bialgebra $(\g_{L,R},\delta_{L,R})$ is coboundary, we can obtain the Poisson structure directly from the Sklyanin bracket~(\ref{Sklyanin}), which I here rewrite for clarity:
\begin{equation}
\left\lbrace a, b \right\rbrace = \frac{1}{2} r^{ij} \cdot((X_i^R\wedge X^R_j - X_i^L\wedge X^L) (a\ot b)).
\label{Sklyanin2}
\end{equation} 

Parametrizing a generic element of $G_{L,R}$ as (for simplicity I omit the subscript $L$ and $R$ in the rest of this section unless otherwise specified)
\begin{equation}
 g=\exp\left(\tilde{h}\rho\left(H\right)+\tilde{x}_{+}\rho\left(X_{+}\right)+\tilde{x}_{-}\rho\left(X_{-}\right)\right) , 
 \label{parametrizationSU2}
\end{equation} 
we obtain the left and right invariant vector fields defined, for a basis $\{e_i\}$ of $\g$, by the relations 
\begin{equation}
X_{e_i}^{R}f\left(g\right)=\frac{d}{dt}\Big|_{t=0}f\left(e^{-te_i}g\right),\qquad
X_{e_i}^{L}f\left(g\right)=\frac{d}{dt}\Big|_{t=0}f\left(ge^{te_i}\right), 
\end{equation} 
as 
\begin{equation}
\begin{gathered}\begin{split}X_{H}^{L}= & \frac{1}{2}\text{csch}\left(\frac{\lambda}{2}\right)\text{sech}\left(\frac{\lambda}{2}\right)\left(\lambda\cosh\left(\lambda\right)+\frac{1}{4}\left(\frac{\sinh\left(\lambda\right)}{\lambda^{2}}-\frac{\cosh\left(\lambda\right)}{\lambda}\right)h^{2}\right)\partial_{h}\\
 & +\frac{1}{2}\left(-1+\frac{1-\lambda\coth\left(\lambda\right)}{2\lambda^{2}}h\right)x_{+}\partial_{x_{+}}+\frac{1}{2}\left(1+\frac{1-\lambda\coth\left(\lambda\right)}{2\lambda^{2}}h\right)x_{-}\partial_{x_{-}},
\end{split}
\\
\begin{split}X_{X_{+}}^{L}= & \left(\left(-1+\frac{1-\lambda\coth\left(\lambda\right)}{2\lambda^{2}}h\right)x_{-}\right)\partial_{h}\\
 & +\left(\lambda\coth\left(\lambda\right)+\frac{1}{2}h+\frac{1-\lambda\coth\left(\lambda\right)}{2\lambda^{2}}x_{-}x_{+}\right)\partial_{x_{+}}+\left(\frac{1-\lambda\coth\left(\lambda\right)}{2\lambda^{2}}x_{-}^{2}\right)\partial_{x_{-}},
\end{split}
\\
\begin{split}X_{X_{-}}^{L}= & \left(\left(1+\frac{1-\lambda\coth\left(\lambda\right)}{2\lambda^{2}}h\right)x_{+}\right)\partial_{h}\\
 & +\left(\frac{1-\lambda\coth\left(\lambda\right)}{2\lambda^{2}}x_{+}^{2}\right)\partial_{x_{+}}+\left(\lambda\coth\left(\lambda\right)-\frac{1}{2}h+\frac{1-\lambda\coth\left(\lambda\right)}{2\lambda^{2}}x_{-}x_{+}\right)\partial_{x_{-}},
\end{split}
\end{gathered}
\end{equation} 
\begin{equation}
\begin{gathered}\begin{split}X_{H}^{R}= & -\frac{1}{2}\text{csch}\left(\frac{\lambda}{2}\right)\text{sech}\left(\frac{\lambda}{2}\right)\left(\lambda\cosh\left(\lambda\right)+\frac{1}{4}h^{2}\left(\frac{\sinh\left(\lambda\right)}{\lambda^{2}}-\frac{\cosh\left(\lambda\right)}{\lambda}\right)\right)\partial_{h}\\
 & -\frac{1}{2}\left(1+\frac{1-\lambda\coth\left(\lambda\right)}{2\lambda^{2}}h\right)x_{+}\partial_{x_{+}}-\frac{1}{2}\left(-1+\frac{1-\lambda\coth\left(\lambda\right)}{2\lambda^{2}}h\right)x_{-}\partial_{x_{-}}
\end{split}
\\
\begin{split}X_{X_{+}}^{R}=- & \left(\left(1+\frac{1-\lambda\coth\left(\lambda\right)}{2\lambda^{2}}h\right)x_{-}\right)\partial_{h}\\
 & -\left(\lambda\coth\left(\lambda\right)-\frac{1}{2}h+\frac{1-\lambda\coth\left(\lambda\right)}{2\lambda^{2}}x_{-}x_{+}\right)\partial_{x_{+}}-\left(\frac{1-\lambda\coth\left(\lambda\right)}{2\lambda^{2}}x_{-}^{2}\right)\partial_{x_{-}}
\end{split}
\\
\begin{split}X_{X_{-}}^{R}= & -\left(\left(-1+\frac{1-\lambda\coth\left(\lambda\right)}{2\lambda^{2}}h\right)x_{+}\right)\partial_{h}\\
 & -\left(\frac{1-\lambda\coth\left(\lambda\right)}{2\lambda^{2}}x_{+}^{2}\right)\partial_{x_{+}}-\left(\lambda\coth\left(\lambda\right)+\frac{1}{2}h+\frac{1-\lambda\coth\left(\lambda\right)}{2\lambda^{2}}x_{-}x_{+}\right)\partial_{x_{-}}
\end{split}
\end{gathered}
\end{equation} 

Substituting these vector fields together with the components $r=r^{ij}e_i\ot e_j$ of the $r$-matrix~(\ref{rMatrixSU2}) in~(\ref{Sklyanin2}) we obtain
\begin{equation}
\left\{ \tilde{h},\tilde{x}_{\pm}\right\} =-z\tilde{x}_{\pm},\qquad\left\{ \tilde{x}_{+},\tilde{x}_{-}\right\} =0 
\label{PoissonSU2dual}
\end{equation} 
Notice that, as a consequence of the choice of parametrization~(\ref{parametrizationSU2}), the Poisson brackets~(\ref{PoissonSU2dual}) coincide exactly with the Lie brackets~(\ref{LieAlgebraSU2dual}) of the dual algebra $\g^*$, and not just to linear order in the generators as it is true for any parametrization.

We can trace back the Poisson brackets between the coordinates ($t,x^a,\theta,\xi^a$) associated (``dual'') to the physical symmetry generators $(E,P_a,J,N_a)$, considering the group element of de Sitter
\begin{equation}
g_{dS}= \exp({t\rho{E}+x^a\rho{P}_a+\theta \rho{J}+\xi^a\rho{N}_a})=g_Lg_R .
\end{equation} 
Combining the maps~(\ref{mapLorSU2param}), (\ref{mapCartanSU2param}) and~(\ref{mapPhysLor}) one obtains
\begin{equation}
\begin{gathered}
t=\frac{1}{2\sqrt{\Lambda}}\left(\tilde{h}^{L}-\tilde{h}^{R}\right),\qquad
\theta=\frac{i}{2}\left(\tilde{h}^{L}+\tilde{h}^{R}\right),\\
x^{1}=\frac{1}{2\sqrt{\Lambda}}\left(\tilde{x}_{+}^{L}-\tilde{x}_{-}^{L}+\tilde{x}_{+}^{R}-\tilde{x}_{-}^{R}\right),\\
x^{2}=\frac{i}{2\sqrt{\Lambda}}\left(\tilde{x}_{+}^{L}+\tilde{x}_{-}^{L}+\tilde{x}_{+}^{R}+\tilde{x}_{-}^{R}\right),\\
\xi^{1}=\frac{1}{2}\left(\tilde{x}_{+}^{L}+\tilde{x}_{-}^{L}-\tilde{x}_{+}^{R}-\tilde{x}_{-}^{R}\right),\\
\xi^{2}=\frac{i}{2}\left(\tilde{x}_{+}^{L}-\tilde{x}_{-}^{L}-\tilde{x}_{+}^{R}+\tilde{x}_{-}^{R}\right).
\end{gathered}
\end{equation} 
Using these relations and~(\ref{PoissonSU2dual}) we find
\begin{equation}
\begin{gathered}
\left\{ t,x^{a}\right\} =-\frac{1}{\kappa}x^{a}, \qquad
\left\{ t,\xi^{a}\right\} =-\frac{1}{\kappa}\xi^{a},\\
\left\{ \theta,x^{a}\right\} =-\frac{1}{\kappa}\epsilon^a_{\ b}\xi^{b},\qquad
\left\{ \theta,\xi^{a}\right\} =-\frac{\Lambda}{\kappa}\epsilon^a_{\ b}x^{b}\\
\left\{ \theta,t\right\} =\left\{ \xi^{a},\xi^{b}\right\} =\left\{ x^{a},x^{b}\right\} =\left\{ \xi^{a},x^{b}\right\} =0.
\end{gathered}
\end{equation} 

We see that the Poisson brackets are linear in the coordinates exactly. It follows that the quantization can be performed trivially by substituting the Poisson brackets with commutators promoting the coordinates to non-commutative coordinates
\begin{equation}
\begin{gathered}
\left[ \hat{t},\hat{x}^{a}\right] =-\frac{1}{\kappa}\hat{x}^{a}, \qquad
\left[ \hat{t},\hat{\xi}^{a}\right] =-\frac{1}{\kappa}\hat{\xi}^{a},\\
\left[ \hat{\theta},\hat{x}^{a}\right] =-\frac{1}{\kappa}\epsilon^a_{\ b}\hat{\xi}^{b},\qquad
\left[ \hat{\theta},\hat{\xi}^{a}\right] =-\frac{\Lambda}{\kappa}\epsilon^a_{\ b}\hat{x}^{b}\\
\left[ \hat{\theta},\hat{t}\right] =\left[ \hat{\xi}^{a},\hat{\xi}^{b}\right] =\left[ \hat{x}^{a},\hat{x}^{b}\right] =\left[ \hat{\xi}^{a},\hat{x}^{b}\right] =0.
\end{gathered}
\label{NCcoordinates}
\end{equation} 
These relations can be understood as the ones defining the non-commutative spacetime on which the $\kappa$-de Sitter symmetries defined in the previous section act covariantly.
The subset of commutation between the coordinates $(\hat{t},\hat{x}^a)$ ``dual'' to the translation generators $(E,P_a)$ can be identified with the ones of $\kappa$-Minkowski spacetime
\begin{equation}
\left[ \hat{t},\hat{x}^{a}\right] =-\frac{1}{\kappa}\hat{x}^{a}, \qquad
\left[ \hat{x}^{a},\hat{x}^{b}\right] = 0.
\end{equation} 
However we see that we have also non-vanishing commutators between $\hat{\theta}$ and $\hat{\xi}^a$ that depend on the cosmological constant and are peculiar of the spacetime associated to $\kappa$-de Sitter.

\section{Contraction to $\kappa$-Poincaré and $\kappa$-Minkowski and bicrossproduct basis}
\label{sec:contraction}

In this last section I show how the contraction limit for vanishing cosmological constant is well defined, and leads to $\kappa$-Poincaré~\cite{Lukierski:1991pn,MajidRuegg}.
Taking the limit $\Lambda \rightarrow 0$ in Eqs.~(\ref{kdeSitterAlg}) and~(\ref{kdeSitterCopr}), we find
\begin{equation}
\begin{gathered}\left[E,P_{a}\right]=0,\qquad\left[P_{1},P_{2}\right]=0,\qquad\left[N_{a},E\right]=-P_{a},\\
\left[N_{a},P_{b}\right]=-\delta_{ab}\kappa\sinh\left(E/\kappa\right),\qquad\left[N_{1},N_{2}\right]=-M\cosh\left(E/\kappa\right),\\
\left[M,N_{a}\right]=\epsilon_{a}^{\ b}N_{b},\qquad\left[M,P_{a}\right]=\epsilon_{a}^{\ b}P_{b},\qquad\left[M,E\right]=0.
\end{gathered} 
\label{kPoincStandAlg}
\end{equation} 
\begin{equation}
\begin{gathered}\Delta E=E\otimes1+1\otimes E\ ,\qquad\Delta M=M\otimes1+1\otimes M\ ,\\
\Delta P_{a}=P_{a}\otimes e^{\frac{1}{2}E/\kappa}+e^{-\frac{1}{2}E/\kappa}\otimes P_{a},\\
\Delta N_{a}=N_{a}\otimes e^{\frac{1}{2}E/\kappa}+e^{-\frac{1}{2}E/\kappa}\otimes N_{a}-\frac{1}{2\kappa}\epsilon_{ab}\,\left(P_{b}\otimes e^{\frac{1}{2}E/\kappa}M-e^{-\frac{1}{2}E/\kappa}M\otimes P_{b}\right)\,.
\end{gathered}
\label{kPoincStandCopr}
\end{equation} 
It is important to stress how, in order to have a convergent limit in the contraction $\Lambda\rightarrow 0$, it is crucial that the deformation parameters $z_L$ and $z_R$ for the two $U_q(\su(2))$ copies ($q_{L,R}=\exp{z_{L,R}}$) have opposite sign, a feature that was also noticed  in~\cite{Cianfrani:2016ogm} in a different quantization scenario for the Euclidean case.

The non-commutative coordinates become, in the limit $\Lambda\rightarrow 0$,
\begin{equation}
\begin{gathered}
\left[ \hat{t},\hat{x}^{a}\right] =-\frac{1}{\kappa}\hat{x}^{a}, \qquad
\left[ \hat{t},\hat{\xi}^{a}\right] =-\frac{1}{\kappa}\hat{\xi}^{a}, \qquad
\left[ \hat{\theta},\hat{x}^{a}\right] =-\frac{1}{\kappa}\epsilon^a_{\ b}\hat{\xi}^{b}, \\
\left[ \hat{\theta},\hat{t}\right] =\left[ \hat{\xi}^{a},\hat{\xi}^{b}\right] =\left[ \hat{x}^{a},\hat{x}^{b}\right] =\left[ \hat{\xi}^{a},\hat{x}^{b}\right] =
\left[ \hat{\theta},\hat{\xi}^{a}\right] =0.
\end{gathered}
\end{equation} 

The algebra~(\ref{kPoincStandAlg}) (\ref{kPoincStandCopr}) is the $\kappa$-Poincaré algebra in standard basis~\cite{Lukierski:1991pn}. It is well known that it is possible to redefine the generators in order to obtain the so-called bicrossproduct basis~\cite{MajidRuegg}. The change of basis can be performed (see for instance~\cite{Ball3D2004} for the AdS case) at the $k$-de Sitter level, before taking the limit $\Lambda\rightarrow 0$, through the maps
\begin{equation}
\begin{gathered}\tilde{E}=E,\qquad\tilde{M}=M,\\
\tilde{P}_{a}=e^{-\frac{1}{2\kappa}E}\left(\cos\left(\frac{\sqrt{\Lambda}}{2\kappa}M\right)P_{a}-\epsilon_{ab}\sqrt{\Lambda}\sin\left(\frac{\sqrt{\Lambda}}{2\kappa}M\right)N_{b}\right),\\
\tilde{N}_{a}=e^{-\frac{1}{2\kappa}E}\left(\cos\left(\frac{\sqrt{\Lambda}}{2\kappa}M\right)N_{a}-\epsilon_{ab}\frac{1}{\sqrt{\Lambda}}\sin\left(\frac{\sqrt{\Lambda}}{2\kappa}M\right)P_{b}\right).
\end{gathered} 
\end{equation} 
We find the algebra
\begin{equation}
\begin{gathered}\left[\tilde{E},\tilde{P}_{a}\right]=\Lambda\tilde{N}_a,\qquad\left[\tilde{N}_{a},\tilde{E}\right]=-\tilde{P}_{a},\\
\left[\tilde{P}_{1},\tilde{P}_{2}\right]=\frac{\Lambda}{2}\frac{\sin\left(2\sqrt{\Lambda}\tilde{M}/\kappa\right)}{\sinh\left(\sqrt{\Lambda}/\kappa\right)},\qquad\left[\tilde{N}_{1},\tilde{N}_{2}\right]=-\frac{1}{2}\frac{\sin\left(2\sqrt{\Lambda}\tilde{M}/\kappa\right)}{\sinh\left(\sqrt{\Lambda}/\kappa\right)},\\
\begin{split}\left[\tilde{N}_{a},\tilde{P}_{b}\right]= & -\frac{\delta_{ab}\sqrt{\Lambda}}{\sinh(\frac{\sqrt{\Lambda}}{\kappa})}\left(\frac{1-e^{-2\tilde{E}/\kappa}}{2}-\sin^{2}\left(\frac{\sqrt{\Lambda}}{\kappa}\tilde{M}\right)\right)\\
 & -\frac{\delta_{ab}}{2\kappa}\left(\mathbf{\tilde{P}}^{2}-\Lambda\mathbf{\tilde{N}}^{2}\right)+\frac{1}{\kappa}\left(\tilde{P}_{a}\tilde{P}_{b}-\Lambda\tilde{N}_{a}\tilde{N}_{b}\right),
\end{split}
\\
\left[\tilde{M},\tilde{N}_{a}\right]=\epsilon_{a}^{\ b}\tilde{N}_{b},\qquad\left[\tilde{M},\tilde{P}_{a}\right]=\epsilon_{a}^{\ b}\tilde{P}_{b},\qquad\left[\tilde{M},\tilde{E}\right]=0.
\end{gathered} 
\end{equation} 
and coalgebra
\begin{equation}
\begin{gathered}\Delta\tilde{E}=\tilde{E}\otimes1+1\otimes\tilde{E}\ ,\qquad\Delta\tilde{M}=\tilde{M}\otimes1+1\otimes\tilde{M}\ ,\\
\Delta\tilde{P}_{a}=e^{-\tilde{E}/\kappa}\otimes\tilde{P}_{a}+\tilde{P}_{a}\otimes\cos\left(\frac{\sqrt{\Lambda}}{\kappa}\tilde{M}\right)-\sqrt{\Lambda}\epsilon_{ab}\tilde{N}_{b}\otimes\sin\left(\frac{\sqrt{\Lambda}}{\kappa}\tilde{M}\right)\\
\Delta\tilde{N}_{a}=e^{-\tilde{E}/\kappa}\otimes\tilde{N}_{a}+\tilde{N}_{a}\otimes\cos\left(\frac{\sqrt{\Lambda}}{\kappa}\tilde{M}\right)-\sqrt{\Lambda}\epsilon_{ab}\tilde{P}_{b}\otimes\sin\left(\frac{\sqrt{\Lambda}}{\kappa}\tilde{M}\right)
\end{gathered}
\end{equation} 
which reduces to the standard $\kappa$-Poincaré Hopf algebra in bicrossproduct basis~\cite{MajidRuegg} for $\Lambda\rightarrow 0$.

\section{Conclusions}

In this paper I have shown how the deformation-quantization of the Chern-Simons action for three dimensional Lorentzian gravity with cosmological constant coupled to a point particle leads to relativistic symmetries of $\kappa$-de Sitter type.
These are defined as the Hopf-algebra of symmetry generators which tends, in the limit of vanishing $\Lambda$, to the three dimensional version of the $\kappa$-Poincar\'e symmetries~\cite{Lukierski:1991pn,MajidRuegg} both in standard and bicrossproduct basis.

This result seems to contradict some previous observation~\cite{SchrMeuskPoinc} asserting that the $\kappa$-Poincar\'e symmetries are not compatible with 3D gravity.
The difference respect to previous approaches resides in the implementation of a new $r$-matrix encoding the deformation-quantization, compatible with the scalar product at the basis of the Chern-Simons action.

In order to obtain the result presented in this paper I took advantage of a splitting of the three dimensional de Sitter algebra in terms of two mutually commuting $\su(2)$, and I applied the methods proposed in~\cite{BallMethod}, based on the ``quantum duality principle''.

I obtain moreover the non-commutative spacetime associated to this set of deformed symmetries.
Having at disposal both the set of symmetries and the defining spacetime commutators, it would be interesting to study the kinematical implications of the construction here presented for a particle living in such a three dimensional scenario.
This would provide a toy-model on which to test some of the implications of $\kappa$-deformed relativistic symmetries with non-vanishing cosmological constant.

At the same time it would be worth investigating the possibility of generalizing the results here presented to the four dimensional case. Even if a Chern-Symons formulation of 4D gravity is not available, the construction of mutually dual Poisson-Lie groups is possible, as for instance shown in~\cite{Ballesteros:2016bml}. It would be interesting to perform a similar construction for a 4D generalization of the r-matrix proposed in this manuscript.

\section*{Aknowledgements}

I would like to thank Prof. Jerzy Kowalski-Glikman and Dott. Stefano Bianco for the helpful discussions during the writing of this manuscript.
This work was supported by funds provided by the National Science Center under
the agreement DEC-2011/02/A/ST2/00294.

\end{document}